# Topologically Induced Optical Activity in Graphene-Based Meta-Structures

*Dmitry A. Kuzmin[†, ‡ *], Igor V. Bychkov[†, ‡], Vladimir G. Shavrov[§], Vasily V. Temnov[¶, ¶¶].*

[†]Chelyabinsk State University, 129 Br. Kashirinykh Str., Chelyabinsk 454001, Russian Federation

[‡]South Ural State University (National Research University), 76 Lenin Prospekt, Chelyabinsk 454080, Russian Federation

[§]Kotelnikov Institute of Radio-engeneering and Electronics of RAS, 11/7 Mokhovaya Str., Moscow 125009, Russian Federation

[¶]Institut des Molécules et Matériaux du Mans, CNRS UMR 6283, Université du Maine, 72085 Le Mans cedex, France

[¶¶]Groupe d'Etude de la Matiere Condensée (GEMaC), Université de Versailles-Saint Quentin en Yvelines, CNRS UMR 8635, Université Paris-Sacley, 45 avenue des Etats-Unis, 78035 Versailles Cedex, France

ABSTRACT. Non-reciprocity and asymmetric transmission in optical and plasmonic systems is a key element for engineering the one-way propagation structures for light manipulation. Here we investigate topological nanostructures covered with graphene-based meta-surfaces, which consist of a periodic pattern of sub-wavelength stripes of graphene winding around the (meta-) tube or (meta-)torus. We establish the relation between the topological and plasmonic properties in these structures, as justified by simple theoretical expressions. Our results demonstrate how to use strong asymmetric and chiral plasmonic responses to tailor the electrodynamic properties in topological meta-structures. Cavity resonances formed by elliptical and hyperbolic plasmons in meta-structures are sensitive to the one-way propagation regime in a finite length (Fabry-Perot-like) meta-tube and display the giant mode splitting in a (Mach-Zehnder-like) meta-torus.

In recent years, enormous attention of researchers has been paid to metamaterials and surfaces, the artificial periodic arrangements of sub-wavelength size elements (so-called "meta-atoms").[1-4] Recently, a piece of a relatively thin nanostructured hyperbolic metamaterial has been used to observe the phenomenon of strongly asymmetric optical transmission in the visible frequency range[5]. A limiting case of ultrathin meta-surfaces for light manipulation is interesting in the view of anomalous reflection,[6] diffraction-free propagation,[7] generation of optical vortexes,[8] the photonic spin Hall effect,[9] etc.

Reconfigurable meta-surfaces may be created on the basis of graphene (the one-atom-thin honey-comb-like carbon lattice).[10, 11] Such meta-surfaces are constructed by densely-packed graphene strips, allow for the electrical control its topology from elliptic to hyperbolic through highly anisotropic σ-near-zero properties, what has a great interest for guiding plasmons manipulation.

For real applications, only metasurfaces with finite size may be used. Presence of the structure edges leads to undesirable losses due to electromagnetic radiation into the outer medium. Such disadvantage may be avoided in cylindrical structures. Graphene-based cylindrical waveguides may operate in single- and multi- mode regimes in frequency range from THz to mid-IR.[12-14] They may support TE- polarized plasmons,[15] similarly to the single graphene layer[16]. Recently, we have shown that cylindrical graphene-based waveguide filled by gyrotropic (or magnetized) medium demonstrate the giant Faraday rotation of high-order plasmonic modes spiraling around the nanowire axis.[17] Although the magnetic control in such hybrid magneto-plasmonic structures is possible, for practical applications it would be preferential to develop similar functionalities without the use of the external magnetic field. In this Letter we focus on the symmetry breaking of chiral SPPs propagating on chiral cylindrical plasmonic waveguides based on rolled graphene metasurfaces (meta-tubes) as well as the finite-length structures serving as meta-cavities.

The main concept of this study is shown in Figure 1. Chiral, azimuthal plasmonic modes propagating along the cylindrical structures are somewhat analogous to the nuts on the screws. Higher order plasmonic modes possess $2m$ nodes, giving the angular intensity distribution visually resembling the shape of common nuts. Whereas in a mechanical case the rotation direction of the nut is determined by the (left- or right-handed) thread on the screw, in plasmonics both rotation directions are generally possible giving rise to the propagating electromagnetic modes rotating clock (+m) or counterclockwise (-m). However, under appropriate conditions, in analogy to the mechanical nut-on-the-screw example, the chirality of the propagating plasmonic modes is dictated by the chirality of the meta-tube and modes with the opposite chirality cannot propagate.

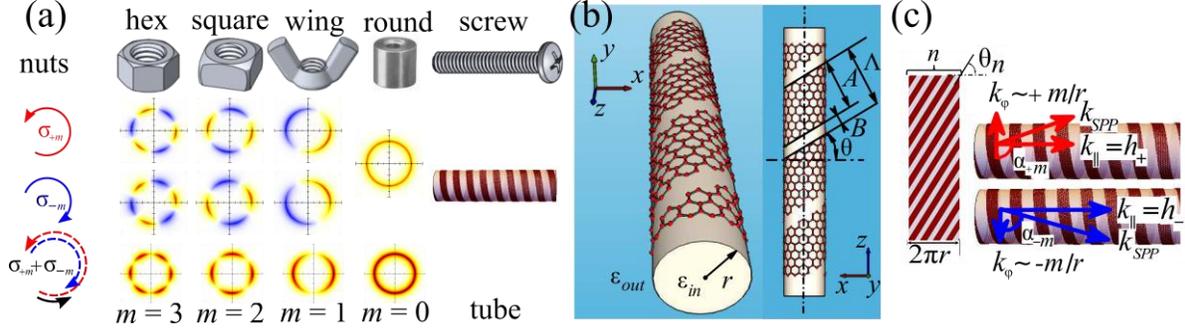

**Figure 1.** (a) Chiral SPP modes are characterized by azimuthal indices $\pm m$ and propagate along plasmonic nanowires similar nuts on a screw. Arrows show the direction of rotation of SPP field distribution for $+m$ and $-m$ modes upon propagation. (b) A chiral graphene-based meta-tube (plasmonic screw) is obtained by winding a bunch of $n$ identical graphene stripes around the cylindrical core under the fixed angle $\theta_n$. (c) Chiral SPP modes with opposite azimuthal numbers $+m$ and $-m$ propagate along the chiral meta-tube with different k-vectors $h_+ \neq h_-$. The intensity distribution of the superposition of $+m$ and $-m$ modes in (a) also rotates upon propagation due to the difference in wavevectors $h_+$ and $h_-$ (see text for details).

Let us consider a dielectric cylinder (core of the waveguide) with dielectric permittivity $\varepsilon_{in} = \varepsilon^r_{in}\varepsilon_0$ (we use SI units, $\varepsilon_0$ is electric constant) and radius $r$, which is coiled by graphene stripe (see Figure 1(b,c)). Such cylinder is embedded in the dielectric medium with dielectric permittivity $\varepsilon_{out} = \varepsilon^r_{out}\varepsilon_0$. Both mediums are non-magnetic ($\mu_{in} = \mu_{out} = \mu_0$) and we use cylindrical coordinates ($\rho$, $\varphi$, $z$), where the $z$-axis coincides with the cylinder axis.

The topographic map projection of our cylindrical structure of radius $r$ in Figure 1(b) is a meta-surface formed by graphene stripes with the width $A$ separated by the spacer width $B$. For a fixed periodicity of the meta-surface $\Lambda = A + B$, the tilt angle can possess the discrete values (see Figure 1(c))

$$\theta_n = \arcsin[nL / 2\pi r] \qquad (1)$$

Here $n$ is an integer, to be denoted as the "topological index", which is the number of graphene stripes winding around the meta-tube. It represents the topological index of the structure because under homeomorphic transformations one cannot change the count of the spirals. The maximum topological index $n_{max} = 2\pi r/\Lambda$ corresponds to the longitudinal orientation of graphene stripes. Similar systems with multiple metallic helices to generate near-fields with high optical chirality have been proposed recently.[18]

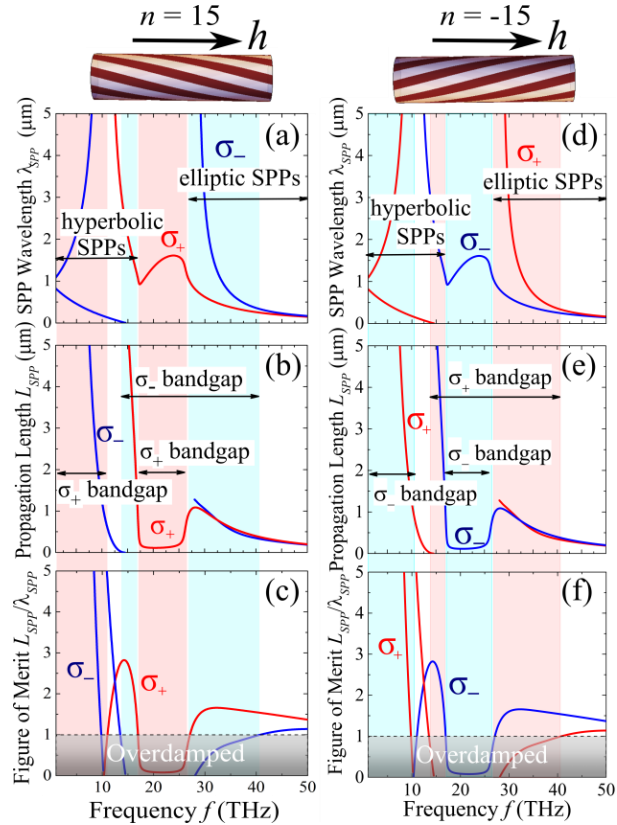

**Figure 2.** Dispersion of SPP modes depends both on the chirality of the meta-tube (index $n$=+15, right-handed tube or $n$=-15, left-handed tube) and the chirality of the mode (index $m = \pm 1$ for $\sigma_\pm$ - modes). Hyperbolic and elliptic SPP modes are separated by bandgaps, where SPPs cannot propagate. The latter are defined in the figure of merit as frequency regions with $L_{SPP}/\lambda_{SPP} < 1$ (overdamped regions, below the dashed horizontal line). Different bandgaps for $\sigma_+$ and $\sigma_-$ modes lead to the phenomenon of one-way propagation; SPP dispersion is symmetric upon the change of tube chirality: $\sigma_+ \rightarrow \sigma_-$ upon $n \rightarrow -n$.

SPPs propagating along the cylindrical meta-tube are described by electric and magnetic fields $\mathbf{E}$, $\mathbf{H} \sim \exp[-i\omega t + ihz + im\varphi]$, where $\omega$ is the circular frequency, $h$ is the propagation constant, $m$ is the azimuthal mode index characterizing SPP's chirality. These azimuthal modes can be interpreted as plane electromagnetic waves characterized by the longitudinal and transversal (to the nanowire axis) components of the wave vector $h_\pm$ and $k_\varphi \approx \pm m/r$, respectively. Two modes with opposite $\pm m$ propagate at different angles with respect to the graphene stripes in our chiral structure. Calculations show that propagation constants for these modes are different, $h_+ \neq h_-$, similarly to the plasmonic modes in gyrotropic graphene-covered nanowires.[17] We will focus on the modes with $m = \pm 1$ (to be denoted as $\sigma_\pm$) and discuss their dispersion characteristics in detail.

For sub-wavelength periodicity, $\Lambda << \lambda$, the optical properties of graphene meta-surfaces are determined by the highly anisotropic conductivity tensor

$$\hat{\sigma}_{meta} = \begin{pmatrix} \sigma_{\varphi\varphi} & \sigma_{\varphi z} \\ \sigma_{z\varphi} & \sigma_{zz} \end{pmatrix}, \qquad (2)$$

where all tensor components depend on graphene conductivity $\sigma_g$ and the capacitive coupling $\sigma_C$ between the stripes (see the Supporting Information for details). Graphene meta-surfaces display the transition from the elliptic to the hyperbolic topology around σ-near-zero case determined by condition of $\text{Im}\{A\sigma_C + B\sigma_g\} = 0$.[10, 11] This transition plays a crucial role in plasmonics: whereas in case of elliptic topology, SPPs can propagate in all directions, in hyperbolic case their propagation is allowed is some specific directions only. Physically, the hyperbolic meta-surface displays the metal-like behavior in one direction while showing the dielectric-like response in the orthogonal directions. Due to the pronounced frequency dependence of $\sigma_g$ and $\sigma_C$ the spectral regions of hyperbolic and elliptic topology are separated by the bandgap: an effect that we are going to use later in the manuscript.

Figure 2 shows the frequency dependence of SPP wavelength $\lambda_{SPP} = 2\pi/\text{Re}\{h\}$, propagation length $L_{SPP} = 1/(2\text{Im}\{h\})$ and the figure of merit $L_{SPP}/\lambda_{SPP}$ for a structure with the core radius $r = 200$ nm, graphene strips of the width $A = 45$ nm, the periodicity $\Lambda = 2\pi r/n_{\max} \approx 50.3$ nm, graphene chemical potential $\mu_{ch} = 0.5$ eV, $\varepsilon^r_{in} = 3$, $\varepsilon^r_{out} = 1$, and the topological index $n = \pm 15$ (or, equivalently, $\theta_n \approx \pm 37^\text{o}$). In chiral structures the wave vectors of $\sigma_+$ ($m$=1) and $\sigma_-$ ($m$ = -1) SPP-modes are oriented differently with respect to graphene stripes (see Figure 1(c)). This leads to the difference in their dispersion relation: whereas $\sigma_+$ SPPs possess a cut-off frequency of 10 THz for $n = 15$, as in case of continuous graphene coating,[11, 12] $\sigma_-$ SPPs are allowed to propagate at lower frequencies. In other words, $\sigma_+$ and $\sigma_-$ dispersion curves are characterized by slightly different bandgaps where SPPs cannot propagate. Change in the structure chirality leads to the opposite behavior. The bandgap opening is caused by the transition from the elliptic (above the bandgap) to the hyperbolic (below the bandgap) SPPs dispersion. This transition occurs at highly anisotropic σ-near-zero points of the metasurface, where their resonant response is accompanied by large dissipative losses. We will show below that the difference in bandgaps can be used to design structures for the asymmetric one-way SPP propagation.

In order to highlight the importance of $\sigma_\pm$ SPP-modes here we note that their excitation has readily been observed in the experiments on isotropic nanowires excited at their tips by a plane electromagnetic wave under oblique incidence.[19-21] In this non-chiral case a helical rotation of intensity distribution was formed by two interfering SPP modes, i.e. m=0 and $\sigma_+$ or $\sigma_-$ modes.[19] In chiral structures we expect to observe the rotation of electromagnetic field distribution solely with a superposition between the distinct $\sigma_+$ and $\sigma_-$ eigenmodes because of their different propagation constants $h_+ \neq h_-$ . Assuming that a linearly polarized electromagnetic wave impinging on the tip of our meta-tube at $z=0$ will predominantly excite the linear combination of $\sigma_+$ and $\sigma_-$ SPPs with equal amplitudes, the resulting azimuthal field distribution will rotate upon propagation as shown in the inset of Fig. 3. Given the case that the difference in the attenuation of $\sigma_+$ and $\sigma_-$ SPPs at $z = z_0$ is equal, i.e. $z_0|\text{Im}\{h_-\} - \text{Im}\{h_+\}| << 1$, the initial field distribution will be preserved and rotated around the tube axis by the angle $\psi = z_0(\text{Re}\{h_-\} - \text{Re}\{h_+\})/2$. After propagation along the meta-tube of finite length $z_0$ the SPPs would be out-coupled into the linearly polarized free-space radiation with the polarization plane rotated by the angle $\psi$. To quantify this polarization rotation, we introduce the specific rotation angle $\psi_0 = \psi/z_0$ per unit length. Figure 3 shows the dependence of this specific rotation angle on the topological index $n$ for a meta-tube with the periodicity $\Lambda = 2\pi r/25 \approx 50.3$ nm, and graphene strip width $A = 45$ nm. The maximum specific rotation can reach a few hundreds of degrees per micrometer at $n=20\ldots22$, corresponding to the tilt angles $\theta_n$ ranging from $50^\circ$ to $60^\circ$. Adjusting the graphene's chemical potential, though the external gate voltage or intrinsic chemical doping can significantly modify the specific rotation $\psi_0$. The maximum values of $\psi_0$ should be much larger than those predicted for the gyrotropic graphene-covered nanowires under similar conditions.[17]

An opposite chirality of the structure can be formally obtained either by assuming negative tilt numbers (or tilt angles) or backward propagating waves. It is worth mentioning that the discussed optical activity in spiral meta-structures is analogous to that in natural sugar solutions and artificial chiral media.[22-25]

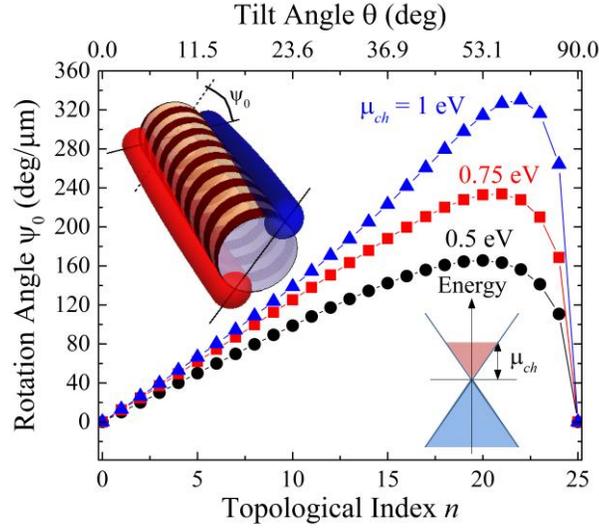

**Figure 3.** Specific rotation angle $\psi_0$ for the spiral waveguide with the periodicity $\Lambda = 2\pi r/25 \approx$ 50.3 nm, and graphene strip width $A = 45$ nm via the topological index of the structure for the frequency 50 THz and different values of the graphene chemical potential. The insets show the definition of the rotation angle, and graphene chemical potential (or Fermi level).

Let us consider now a meta-tube of finite length $L$, where forward and backward propagating SPPs might form Fabry-Perot resonances. Due to the asymmetric SPP propagation the resonant condition reads $L[h_+(\omega_{res}) + h_-(\omega_{res})] = 2\pi M$, where $M$ is an integer number. Keeping in mind SPP dispersion in Figure 2, two types of resonances are possible: with hyperbolic SPPs below the bandgap and elliptic SPPs above the bandgap.

Significant shift of the cut-off frequencies for the forward and backward propagating SPPs may prohibit the existence of some lower-order (small $M$) Fabry-Perot modes in chiral meta-tubes (see Figure 4(a)). This effect is illustrated in Figure 4(b), which shows show the resonant frequencies of a meta-tube with the length $L = 1$ μm under variation of the topological index $n$ (or the tilt angle of the stripes).

One can see that within a certain range of the topological indices Fabry-Perot resonances are absent because of the one way propagation regime. For the higher-order Fabry-Perot resonances (large $M$) this range becomes smaller because $h$ increases and $\alpha_{\sigma\pm}$ both get close to $90^{\mathrm{o}}$ (see Figure 1(c)), when the difference between two polarizations $\sigma_{\pm}$ ($m=\pm1$) is negligible.

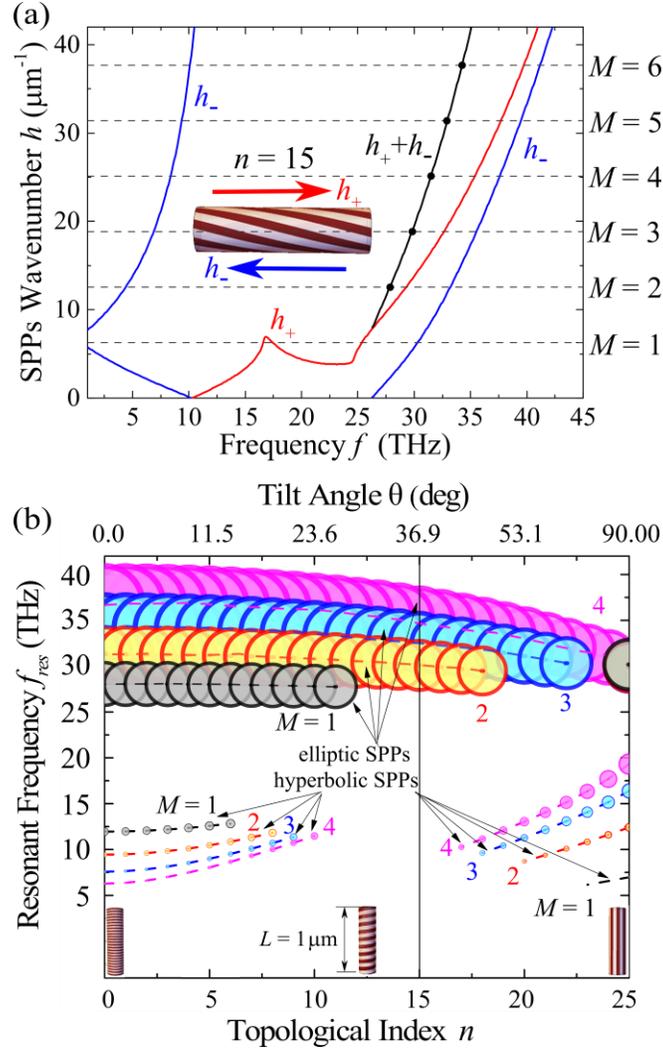

**Figure 4.** (a) Resonant Fabry-Perot modes are obey $[h_+(\omega_{res}) + h_-(\omega_{res})] = 2\pi M/L$ for $M = 1, 2, 3,$ ..., as indicated by the crossing points with dashed horizontal lines $h = 2\pi M/L$ for $L = 1$ μm. SPP dispersion is taken from Fig. 2. (b) Resonant frequencies $f_{res} = \text{Re}[\omega_{res}]/2\pi$ of such a finite-length meta-tube depend on the topological index $n$. The nonexistence of Fabry-Perot resonances within a certain range of topological indices manifests itself in the discontinuity of the curves. Symbol sizes (circle radii) represent the full-width half-maximum of Fabry-Perot resonances.

Excitation of a chiral meta-tube, with parameters corresponding to the one way propagation regime, should lead to the creation of an electromagnetic "hotspot" at one of the tips;[26] for modes with an opposite chirality the hotspot will be located at the opposite tip.

An even more interesting approach to change the topology of the structure is to transform a cylinder into a torus and thus introduce the second topological index $N$ (see Figure 5(a)). In such a system, the tilt angle $\theta_{nN}$ must satisfy two distinct conditions: $\theta_{nN} = \arcsin[n\Lambda/2\pi r]$ and $\theta_{nN} = \arccos[N\Lambda/2\pi R]$. For an arbitrary geometry of the torus these conditions are usually not satisfied for an all $n$. For example, two distinct configurations with perpendicular ($n=0$, $N=N_{max}$) and longitudinal ($n=n_{max}$, $N=0$) orientation of graphene stripes exist only for if the ratio $R/r$ of the torus radii is an integer.

For the analysis of meta-torus resonances we assume $R/r \gg 1$, i.e. we can formally describe it as a piece of the cylinder obeying periodic boundary conditions along the cylinder axis. The Fabry-Perot condition for modes propagating on a torus clock- and counterclockwise read: $2\pi R h_{\pm} = 2\pi M$, where $M$ is an integer number. Given the case that the propagation constants $h_{+}$ and $h_{-}$ are different, the resonant condition for these two modes will be satisfied for two different frequencies. Only in two degenerate cases of $0^{o}$ and $90^{o}$ tilt angle the mode propagation would be symmetric and both resonant frequencies would become identical.

The resonances of all orders exist in the meta-torus for all possible topological indexes in contrast to the finite length meta-tube (see Figure 4(b)).

Let us consider the meta-torus formed by a meta-tube with $r = 200$ nm and $R = 2$ μm, covered by graphene strips with the width $A = 45$ nm, a situation corresponding to $n_{max} = 2\pi r/\Lambda = 25$ and $N_{max} = 2\pi R/\Lambda = 250$. Resonance curves of such a torus are shown in Figure 5(c-f) for several combinations of topological indexes $(n, N) = (0, 250)$, $(15, 200)$, and $(25, 0)$, which correspond to the tilt angles $\theta_{nN} = 0^o$, $37^o$, and $90^o$, respectively.

The resonant modes shown here correspond to the condition $M=1$ and $M=10$. For non-zero chirality of the structure the resonant frequencies for counter- and clockwise propagating modes are different. The resonant condition for the meta-torus $Rh = M$ uniquely defines the effective propagation angle of the SPPs: $\tan(\alpha_{mM}) = Mr/mR$. In analogy to the topological indices $n$ and $N$ a pair of electrodynamic indices $m$ and $M$ defines the electrodynamic topology of the resonant mode. An existence of topological SPP resonances on a meta-torus implies a fixed relation between the structural topological indices $(n, N)$ and electromagnetic topological indices $(m, M)$ of a resonant mode:

$$\frac{n}{N} \operatorname{ctg} \theta_{nN} = \frac{m}{M} \operatorname{tg} \alpha_{mM} \qquad (3)$$

Equation (3) shows that the geometrical chirality of the structure is connected with the electromagnetic chirality of resonant SPP modes. The above analysis facilitates the understanding the splitting of resonant frequencies in Figure 5(c-f), where we again focus on the chiral properties of $\sigma_{\pm}$ ($m = \pm 1$) SPP resonances in structures with longitudinal ($n = 25$, $N = 0$) and perpendicular ($n = 0$, $N = 250$) graphene stripes, as compared to a tilted chiral configuration ($n = 15$, $N = 200$). For the fundamental mode with $m = 0$ there is no splitting for the resonant frequencies of counter-clockwise and clockwise propagation in none of these structures.

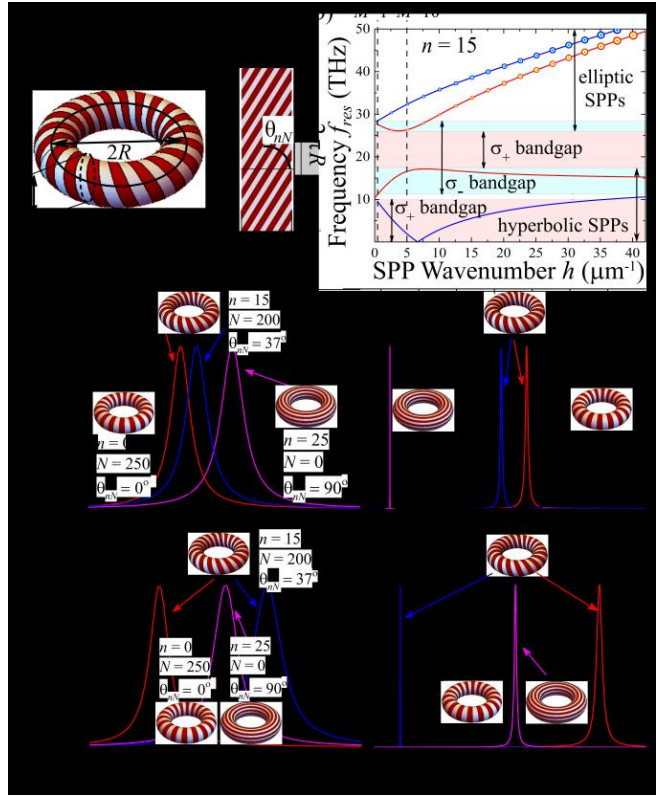

**Figure 5.** (a) A meta-torus is characterized of the pair of topological indices of the structure n and N. (b) Both chiral $\sigma_\pm$ SPP modes exist for all wavenumbers. Resonance curves of the elliptic (c,e) and hyperbolic (d,f) SPPs in the meta-torus correspond to the condition $Rh = M$ for $M = 1$ (c,d) and 10 (e,f); $R/r = 10$. The schematic of the torus is shown near each curve. Continuous red and blue lines correspond to the counter-clockwise and clockwise SPP propagation, respectively.

For $M$=1 in Figure 5(c,d) elliptical and hyperbolic $\sigma_\pm$ modes possess different frequencies for all structures. Figure 5(e) illustrates a specially selected, exotic case of $M$=10 when SPP's propagation angle $\alpha_{\sigma\pm} = 45^\circ$ with respect to graphene stripes is identical for longitudinal and perpendicular stripe orientation. This situation corresponds to the condition of $h_{\sigma\pm} = r^{-1}$ ($= 5$ μm$^{-1}$ for structure in Figure 5). The splitting of resonant elliptic and hyperbolic $\sigma_\pm$ modes for a chiral structure is almost the largest one (see Figure 5(b)).

Surprisingly, Eq. (3) helps to explain why the maximum splitting occurs at the $M$ =10. However, this requires some additional information on the electrodynamics of the structure. The analysis of SPPs propagating on a flat metasurface shows[27] that their frequencies display the largest difference for wavevectors along and perpendicular with respect to graphene stripes, respectively (see the Supporting Information for details). For resonant modes in the chiral toroidal resonator we have equal $h_{\sigma\pm}$, and thus, $\alpha_{\sigma-} = 180^\circ - \alpha_{\sigma+}$. In analogy to planar metasurfaces, the maximum frequency splitting in chiral toroidal structure should be achieved for $\alpha_{\sigma-} - \alpha_{\sigma+} = 90^\circ$, which leads to $\alpha_{\sigma+} = 45^\circ$, $\alpha_{\sigma-} = 135^\circ$; in addition we should have $\theta_{nN} = 45^\circ$. Under these conditions we can use Eq. (3) to calculate the set of topological indices for the maximum frequency splitting obeying $Mn/N = 1$. For the model structure in Figure 4, $\theta_{nN} = 45^\circ$ cannot be achieved for any possible topological indices. The closest angle $\theta_{nN} \approx 37^\circ$ is obtained for $(n,N) = (15, 200)$, resulting in $M = N \cdot \mathrm{tg}(\theta_{nN})/n = 10$.

In conclusion, we have investigated a new class of topological plasmonic structures, which are formed by rolled graphene-based metasurfaces. While a graphene meta-tube displays a giant rotation of azimuthal plasmonic modes, its intrinsic chirality plays a crucial role for the design of "one-way propagation" plasmonic devices and is responsible for the disappearance of Fabry-Perot resonances in finite-length meta-tubes. A piece of meta-tube, rolled in a torus, possesses a distinct spectrum of azimuthal cavity modes with a large splitting for clock- and counterclockwise propagation directions of SPPs. Interestingly, the electromagnetic and geometrical topological indices of the structure are intimately connected by simple analytical expressions, which physical meaning remains to be clarified. Therefore, our results are not limited to graphene-based structures and pave the way for topological plasmonics in chiral plasmonic nanostructures.

ASSOCIATED CONTENT

The following files are available free of charge.

A discussion of theoretical electrodynamics in metasurface-based cylindrical structures as well as the properties of elliptic and hyperbolic SPPs on planar meta-surfaces (PDF)

AUTHOR INFORMATION

**Corresponding Author**

 * kuzminda@csu.ru.

**Author Contributions**

The manuscript was written through contributions of all authors. All authors have given approval to the final version of the manuscript.

**Funding Sources**

The work was financially supported in part by RFBR (16-37-00023, 16-07-00751, 16-29-14045, 17-57-150001), RScF (14-22-00279), Grant of the President of the RF (MK-1653.2017.2), Act 211 Government of the Russian Federation (contract № 02.A03.21.0011), Stratégie internationale NNN-Telecom de la Région Pays de La Loire, and PRC CNRS-RFBR "Acousto-magneto-plasmonics".

**Notes**

The authors declare no competing financial interest.

TOC Figure:

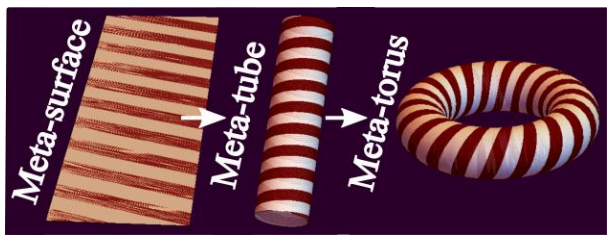